\documentclass[
 twocolumn,
 nofootinbib,
 amsmath,
 amssymb,
 aps,
]{revtex4-2}

\usepackage{graphicx} % Include figure files
\usepackage{dcolumn} % Align table columns on decimal point
\usepackage{bm} % bold math

\usepackage{hyperref} % add hypertext capabilities

\everymath{\displaystyle}

\usepackage{xcolor}
\allowdisplaybreaks
\usepackage{mathtools}
\usepackage[linesnumbered, ruled, vlined]{algorithm2e}

\makeatletter
\newcommand{\vast}{\bBigg@{3.0}}
\newcommand{\Vast}{\bBigg@{4.0}}
\newcommand{\vastt}{\bBigg@{5.0}}

\makeatother

\usepackage{coffeestains}

\usepackage{xr}
\makeatletter
\@addtoreset{equation}{section}
\makeatother

\usepackage{chngcntr}
\makeatletter
\g@addto@macro\appendix{
	\counterwithin{equation}{section}
	
	\counterwithin{figure}{section}
	
}
\makeatother

\begin{document}

\preprint{APS/123-QED}

% \title{Absence of $\mathrm{O} (2)$ symmetry in the original Vicsek model}
\title{Absence of $\mathrm{O} (2)$ symmetry in the Vicsek model}
% \thanks{A footnote to the article title}

\author{Yushin Takahashi}

\author{Kota Mitsui}

\author{Tsuyoshi Mizohata}

\author{Hideyuki Miyahara}
% \author{宮原 英之}

\email{miyahara@ist.hokudai.ac.jp, hmiyahara512@gmail.com}

% \thanks{corresponding author.}

\affiliation{
	Graduate School of Information Science and Technology,
	Hokkaido University, Sapporo, Hokkaido 060-0814, Japan
}

% \affiliation{
%     北海道大学 大学院情報科学研究院 
% }  

% \altaffiliation[Also at ]{Physics Department, XYZ University.}

\date{\today}

\begin{abstract}
	The phase transition in the Vicsek model is widely believed to be associated with spontaneous symmetry breaking of the two-dimensional rotational symmetry $\mathrm{O} (2)$.
	In this paper, we revisit the original Vicsek model introduced by Vicsek \textit{et al.} [Phys.~Rev.~Lett.~75,~1226~(1995)] and demonstrate that the original angle-based update rule is not invariant under global phase shifts at the level of angular increments when it is implemented using the principal-value arctangent, as in the original definition.
	As a consequence, we numerically demonstrate that the phase transition reported in the original paper vanishes when the global phase is adaptively chosen.
\end{abstract}

% \keywords{Suggested keywords}

\maketitle

% \tableofcontents

\section{Introduction}

Active matter refers to a class of nonequilibrium systems composed of many interacting units that consume energy from their surroundings to generate self-propelled motion.
Since the pioneering work of Vicsek~\textit{et al.}~\cite{Vicsek_001}, this field has attracted considerable attention in statistical physics, biophysics, and soft matter physics.
Representative examples include cell motility and collective slime mold dynamics~\cite{Rappel_001}, collective motion in bacterial suspensions~\cite{Wu_003}, the coordinated behavior of fish schools and animal flocks~\cite{Bonabeau_001}, and self-organization in insect societies~\cite{Deneubourg_001, Theraulaz_001}.
A common feature of these systems is that large-scale order and global patterns emerge spontaneously even though each constituent follows only local interaction rules.

In particular, collective motion in systems of self-propelled particles has been extensively studied as a paradigmatic example of how simple interaction rules lead to macroscopic order.
Early experimental and theoretical studies reported the formation of cooperative structures in bacterial and colloidal systems ~\cite{Wu_003, Becco_001}, revealing the universal properties of active matter.
Moreover, the mechanisms of collective decision-making and self-organization in biological populations have been elucidated through studies of ant foraging behavior and swarming~\cite{Deneubourg_001, Theraulaz_001}.

To achieve a unified understanding of such phenomena, theoretical models of self-propelled particles have been actively developed.
Among them, the Vicsek model and its variants~\cite{Vicsek_001, Chate_001, Ramaswamy_001, Bowick_001, Fodor_001, te-Vrugt_001}, which assume only local velocity-alignment interactions, are widely regarded as the simplest theoretical framework that exhibits a phase transition from a disordered state to an ordered state~\cite{Czirok_002}.
This result demonstrates that phase transitions and critical phenomena can arise even in nonequilibrium systems, thereby significantly accelerating the progress of active matter research.

Since the introduction of the Vicsek model, the study of active matter has rapidly expanded, leading to the development of hydrodynamic theories, continuum descriptions, and quantitative comparisons with experiments.
These advances suggest that active matter is not merely a description of biological phenomena but may represent a distinct universality class in nonequilibrium statistical physics.
Indeed, due to the combined effects of energy dissipation and self-propulsion, active matter is now regarded as an archetypal nonequilibrium many-body system exhibiting collective behavior that has no equivalent in equilibrium systems~\cite{Feder_001}.

As mentioned above, the Vicsek model is one of the most important models in active matter~\cite{Vicsek_001, Chate_001, Ramaswamy_001, Bowick_001, Fodor_001, te-Vrugt_001}, and it is widely believed to exhibit a phase transition accompanied by spontaneous symmetry breaking of $\mathrm{O} (2)$ symmetry.
In this paper, we revisit the original definition of the Vicsek model~\cite{Vicsek_001} and examine its properties from the perspective of $\mathrm{O} (2)$ symmetry.
More specifically, we show that the original Vicsek model lacks $\mathrm{O} (2)$ symmetry.
That is, the principal-value formulation of the original Vicsek update rule introduces a branch-cut dependence, which breaks the global phase-shift invariance of the angular dynamics.
For clarity, we refer to this model as the $\arctan$ Vicsek model to distinguish it from another variant considered in this paper.
We further demonstrate that an adaptive choice of the global phase drastically changes the resulting macroscopic behavior.
Notably, this phenomenon is inherent in the original definition of the Vicsek model, which employs trigonometric functions in its update rule.
To support the main claims of this study, we discuss the continuous-time limit of the Vicsek model and the relationship of our findings with Ref.~\cite{Baglietto_001}, where the effect of random rotations of the reference frame on the macroscopic behavior of the Vicsek model was studied, in the Appendixes.

\section{Vicsek model}

In this section, we introduce the Vicsek model proposed in Ref.~\cite{Vicsek_001} and another variant proposed in Ref.~\cite{Miyahara_001}.
To avoid confusion, we call them the $\arctan$ Vicsek model and the arithmetic-mean Vicsek model, respectively, in this paper.
We examine these two models from the perspective of $\mathrm{O} (2)$ symmetry.
As will be clarified later, the two models differ in how the mean direction of the angle variables is defined, and this difference leads to distinct macroscopic flocking behaviors.

The definition of the $\arctan$ Vicsek model is given, for $i = 1, 2, \dots, N$, by~\cite{Vicsek_001}
\begin{subequations} \label{main_eq_definition_original-Vicsek-model_001_001}
	\begin{align}
		\bm{x}_i (t + \Delta t) & = \bm{x}_i (t) + \bm{v}_i (t) \Delta t,                                                                                                             \\
		\bm{v}_i (t)            & = v_\mathrm{abs}
		\begin{bmatrix}
			\cos \theta_i (t) \\
			\sin \theta_i (t)
		\end{bmatrix},                                                                                                                                                             \\
		\theta_i (t + \Delta t) & = \langle \theta_j (t) \rangle_{i, r_\mathrm{V}}^\mathrm{V} + \Xi_i (t + \Delta t, t). \label{main_eq_dynamics-angle_original-Vicsek-model_001_001}
	\end{align}
\end{subequations}
where
\begin{align}
	\langle \theta_j (t) \rangle_{i, r_\mathrm{V}}^\mathrm{V} & \coloneqq \arctan \frac{\langle \sin \theta_j (t) \rangle_{i, r_\mathrm{V}}}{\langle \cos \theta_j (t) \rangle_{i, r_\mathrm{V}}}, \label{main_eq_mean-angle_original-Vicsek-model_001_001}
\end{align}
and
\begin{align}
	\Xi_i (t + \Delta t, t) & \sim \mathrm{uniform} (\cdot; - \eta \sqrt{\Delta t} / 2, \eta \sqrt{\Delta t} / 2).
\end{align}
Here, $\mathrm{uniform} (\cdot; a, b)$ is the uniform distribution over $[a, b]$ and $\| \cdot \|_\mathrm{F}$ is the Frobenius norm for an $n$-dimensional vector $\bm{x} = [x_1, x_2, \dots, x_n]^\intercal$:
\begin{align}
	\| \bm{x} \|_\mathrm{F} & \coloneqq \sqrt{\sum_{i=1}^n x_i^2}.
\end{align}
Note that Eq.~\eqref{main_eq_mean-angle_original-Vicsek-model_001_001} can be transformed into
\begin{align}
	\langle \theta_j (t) \rangle_{i, r_\mathrm{V}}^\mathrm{V} & = \arg (\langle \mathrm{e}^{\mathrm{\mathrm{i} \theta_j (t)}} \rangle_{i, r_\mathrm{V}}), \label{main_eq_mean-angle_original-Vicsek-model_001_002}
\end{align}
where $\arg (\cdot)$ is the argument of a complex number that takes a value in $(-\pi, \pi]$.
The physical interpretations of $\bm{x}_i$, $\bm{v}_i$, and $\theta_i$ in Eq.~\eqref{main_eq_definition_original-Vicsek-model_001_001} are the position, velocity, and direction of the $i$-th agent, respectively.

Next, we consider the following update rule for $\{ \theta_i (t) \}_{i=1}^N$, referred to as the arithmetic-mean Vicsek model~\cite{Miyahara_001} in this paper, instead of Eq.~\eqref{main_eq_dynamics-angle_original-Vicsek-model_001_001}:
\begin{align}
	\theta_i (t + \Delta t) & = \langle \theta_j (t) \rangle_{i, r_\mathrm{V}} + \Xi_i (t + \Delta t, t), \label{main_eq_mean-angle_angle-Vicsek-model_001_001}
\end{align}
and
\begin{align}
	\langle \theta_j (t) \rangle_{i, r_\mathrm{V}} & \coloneqq \frac{1}{N_{i, r_\mathrm{V}} (t)} \sum_{\substack{j = 1, 2, \dots, N : \\ \| \bm{x}_i (t) - \bm{x}_j (t) \|_\mathrm{F} \le r_\mathrm{V}}} \theta_j (t),
\end{align}
where $N_{i, r_\mathrm{V}} (t)$ is the number of neighbors of the $i$-th agent within $r_\mathrm{V}$:
\begin{align}
	N_{i, r_\mathrm{V}} (t) & \coloneqq \# \{ j \in \{ 1, 2, \dots N \} \mid \| \bm{x}_i (t) - \bm{x}_j (t) \|_\mathrm{F} \le r_\mathrm{V} \}.
\end{align}
We note that the difference between the two models is how the mean direction of agents is computed.

To describe the collective behavior of alignment, the order parameter is defined as~\cite{Vicsek_001}
\begin{align}
	v_\mathrm{op} (t) & \coloneqq \frac{1}{v_\mathrm{abs} N} \vast\| \sum_{i = 1}^N \bm{v}_i (t) \vast\|_\mathrm{F}. \label{main_eq_def_order-parameter_Vicsek-model_001_001}
\end{align}

\section{Global phase shift and $\mathrm{O} (2)$ symmetry}

In this section, we revisit the rotational symmetry in phase space and $\mathrm{O} (2)$ symmetry and examine the two models introduced in the previous section from this perspective.
We see that the $\arctan$ Vicsek model, Eq.~\eqref{main_eq_definition_original-Vicsek-model_001_001}, does not possess $\mathrm{O} (2)$ symmetry while the arithmetic-mean Vicsek model, Eq.~\eqref{main_eq_mean-angle_angle-Vicsek-model_001_001}, does.

We consider the following global phase shift for the $\arctan$ Vicsek model, Eq.~\eqref{main_eq_definition_original-Vicsek-model_001_001}:
\begin{align}
	\{ \theta_i (t) \}_{i=1}^N & \mapsto \{ \theta_i (t) + \phi \}_{i=1}^N. \label{main_eq_phase-shift_alpha_Vicsek-model_001_001}
\end{align}
To simplify the discussions, we define the discrete-time difference of $\theta_i (t)$ as follows:
\begin{align}
	\Delta \theta_i (t) & \coloneqq \theta_i (t + \Delta t) - \theta_i (t). \label{main_eq_Delta-theta_Vicsek-model_001_001}
\end{align}
Here, we say that the system of interest is $\mathrm{O} (2)$ symmetric if Eq.~\eqref{main_eq_Delta-theta_Vicsek-model_001_001} is invariant under the global phase shift in Eq.~\eqref{main_eq_phase-shift_alpha_Vicsek-model_001_001}.

From Eqs.~\eqref{main_eq_dynamics-angle_original-Vicsek-model_001_001} and \eqref{main_eq_mean-angle_original-Vicsek-model_001_002}, we get the following form by applying the global phase shift, Eq.~\eqref{main_eq_phase-shift_alpha_Vicsek-model_001_001}, to Eq.~\eqref{main_eq_Delta-theta_Vicsek-model_001_001}:
\begin{align}
	\Delta \theta_i (t) & = \langle \theta_j (t) \rangle_{i, r_\mathrm{V}}^\mathrm{V} + \Xi_i (t + \Delta t, t) - \theta_i (t)                     \\
	                    & = \arg (\langle \mathrm{e}^{\mathrm{i} \theta_j (t)} \rangle_{i, r_\mathrm{V}}) + \Xi_i (t + \Delta t, t) - \theta_i (t) \\
	                    & \mapsto \arg (\langle \mathrm{e}^{\mathrm{i} (\theta_j (t) + \phi)} \rangle_{i, r_\mathrm{V}}) \nonumber                 \\
	                    & \quad + \Xi_i (t + \Delta t, t) - (\theta_i (t) + \phi)                                                                  \\
	                    & = \arg (\langle \mathrm{e}^{\mathrm{i} \theta_j (t)} \rangle_{i, r_\mathrm{V}}) + 2 \pi n (\phi) \nonumber               \\
	                    & \quad + \Xi_i (t + \Delta t, t) - \theta_i (t)                                                                           \\
	                    & = \langle \theta_j (t) \rangle_{i, r_\mathrm{V}}^\mathrm{V} + 2 \pi n (\phi) + \Xi_i (t + \Delta t, t) - \theta_i (t).
\end{align}
where $n$ is determined such that $\phi$ satisfies $\arg (\langle \mathrm{e}^{\mathrm{i} \theta_j (t)} \rangle_{i, r_\mathrm{V}}) + \phi + 2 \pi n (\phi) \in (- \pi, \pi]$.
This additional term, $2 \pi n (\phi)$, is absent only when the branch cut is not crossed.
Hence the angular increment is not in general invariant under the global phase shift, Eq.~\eqref{main_eq_phase-shift_alpha_Vicsek-model_001_001}.

On the other hand, in the case of Eq.~\eqref{main_eq_mean-angle_angle-Vicsek-model_001_001}, we have the following form:
\begin{align}
	\Delta \theta_i (t) & = \langle \theta_j (t) \rangle_{i, r_\mathrm{V}} + \Xi_i (t + \Delta t, t) - \theta_i (t)                       \\
	                    & \mapsto \langle \theta_j (t) + \phi \rangle_{i, r_\mathrm{V}} + \Xi_i (t + \Delta t, t) - (\theta_i (t) + \phi) \\
	                    & = \langle \theta_j (t) \rangle_{i, r_\mathrm{V}} + \Xi_i (t + \Delta t, t) - \theta_i (t).
\end{align}
Obviously, $\mathrm{O} (2)$ symmetry holds in the arithmetic-mean Vicsek model, Eq.~\eqref{main_eq_mean-angle_angle-Vicsek-model_001_001}.

\section{Numerical simulations}

In the previous section, we discussed that $\mathrm{O} (2)$ symmetry is broken in the $\arctan$ Vicsek model, Eq.~\eqref{main_eq_definition_original-Vicsek-model_001_001}.
Here, we demonstrate that the absence of $\mathrm{O} (2)$ symmetry affects the macroscopic behavior of the $\arctan$ Vicsek model, Eq.~\eqref{main_eq_definition_original-Vicsek-model_001_001}.
More specifically, the phase transition reported in Ref.~\cite{Vicsek_001} vanishes completely when we adaptively choose the global phase shift $\phi$ in Eq.~\eqref{main_eq_phase-shift_alpha_Vicsek-model_001_001}.
We also confirm that the arithmetic-mean Vicsek model, Eq.~\eqref{main_eq_mean-angle_angle-Vicsek-model_001_001}, robustly shows the same macroscopic behavior regardless of the choice of the global phase $\phi$ in Eq.~\eqref{main_eq_phase-shift_alpha_Vicsek-model_001_001}.

Setting $\bar{\phi} = 0, \pi$, we consider the following time-dependent global phase shift:
\begin{align}
	\phi (t) & \coloneqq \bar{\phi} - \langle \theta_i (t) \rangle. \label{main_eq_def_phase-factor_001_001}
\end{align}
where
\begin{align}
	\langle \theta_i (t) \rangle & \coloneqq \frac{1}{N} \sum_{i = 1, 2, \dots, N} \theta_i (t).
\end{align}
To remove any ambiguity in the order of operations, we summarize the numerical procedure in Algorithm 1.
In all simulations reported below, the global phase shift is applied to the angular variables before updating the positions and angles at the next time step.
Thus, the shifted angles are used in the subsequent evaluation of the velocities and alignment rule.
\begin{algorithm}[t]
	\DontPrintSemicolon
	\caption{Vicsek model with the global phase shift.}
	\label{main_algo_Vicsek-model-with-phase-shift_005_001}
	% \small
	% \SetAlgoNoLine
	\LinesNumbered
	\Begin{
	initialize $\{ \bm{x}_i (0) \}$ and $\{ \theta_i (0) \}$ by sampling from uniform distributions \;
	compute $\{ \bm{v}_i (0) \}_{i=1}^N$ \;
	\For{$\tau = 1$ \KwTo $N_\tau$}{
	perform the global phase shift: $\{ \theta_i (t) \}_{i=1}^N \mapsto \{ \theta_i (t) + \phi (t) \}_{i=1}^N$ \;
	compute $\{ \bm{x}_i (\tau \Delta t) \}_{i=1}^N$ \;
	compute $\{ \theta_i (\tau \Delta t) \}_{i=1}^N$ \;
	compute $\{ \bm{v}_i (\tau \Delta t) \}_{i=1}^N$ \;
	}
	}
\end{algorithm}
We note that different update conventions have been adopted in the literature.
For example, Ref.~\cite{Baglietto_001} examines whether the ordering behavior of the Vicsek model depends on the order in which the velocities $\{ \bm{v}_i (\tau \Delta t) \}_{i=1}^N$ and the positions $\{ \bm{x}_i (\tau \Delta t) \}_{i=1}^N$ are computed.

\begin{figure}[t]
	\centering
	\includegraphics[scale=0.60]{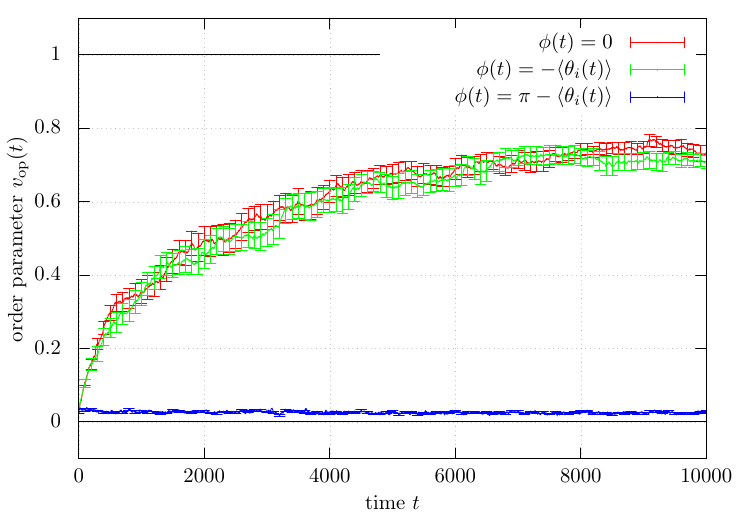}
	\includegraphics[scale=0.60]{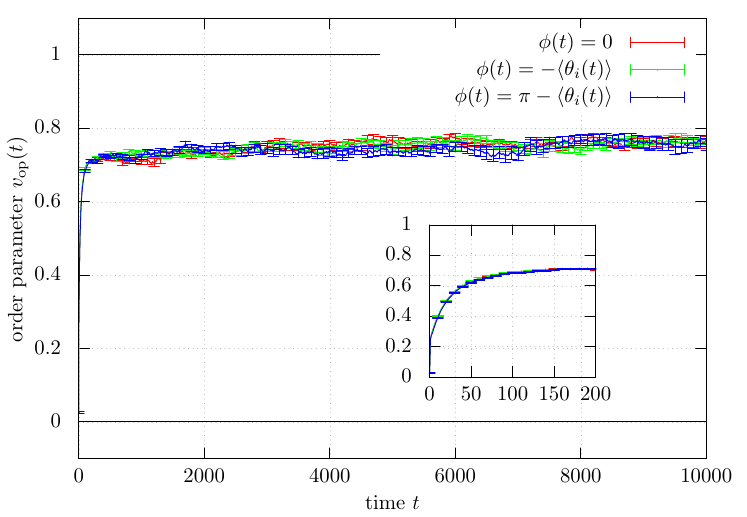}
	\caption{Time evolution of the order parameter, Eq.~\eqref{main_eq_def_order-parameter_Vicsek-model_001_001}, of (upper) the $\arctan$ Vicsek model, Eq.~\eqref{main_eq_definition_original-Vicsek-model_001_001}, and (lower) the arithmetic-mean Vicsek model, Eq.~\eqref{main_eq_mean-angle_angle-Vicsek-model_001_001}. We set $N = 1600$, $L = 8.0$, $v_\mathrm{abs} = 0.01$, $r_\mathrm{V} = 0.1$, $\eta = 0.75$, and $\Delta t = 1.0$. We ran simulations 30 times to compute standard errors. In the inset, we plot the data for $t \in [0, 200]$ to see the relaxation process.}
	\label{main_fig_gnuplot_time-evolution_order-parameter_001_001}
\end{figure}

\begin{figure}[t]
	\centering
	\includegraphics[scale=0.60]{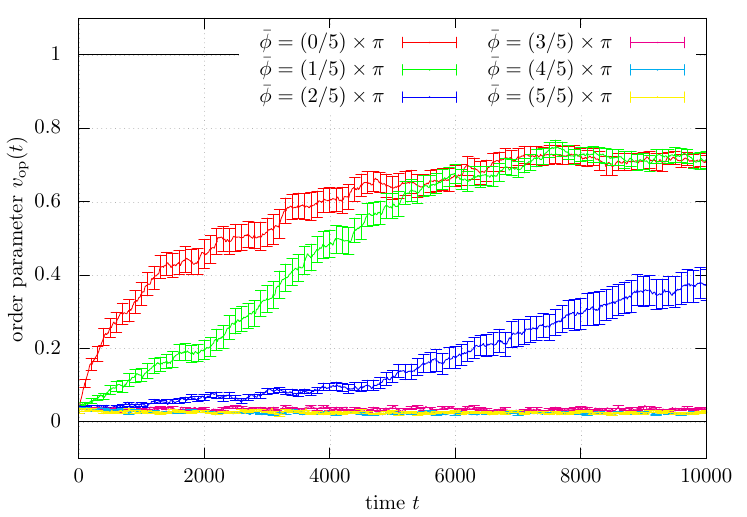}
	\includegraphics[scale=0.60]{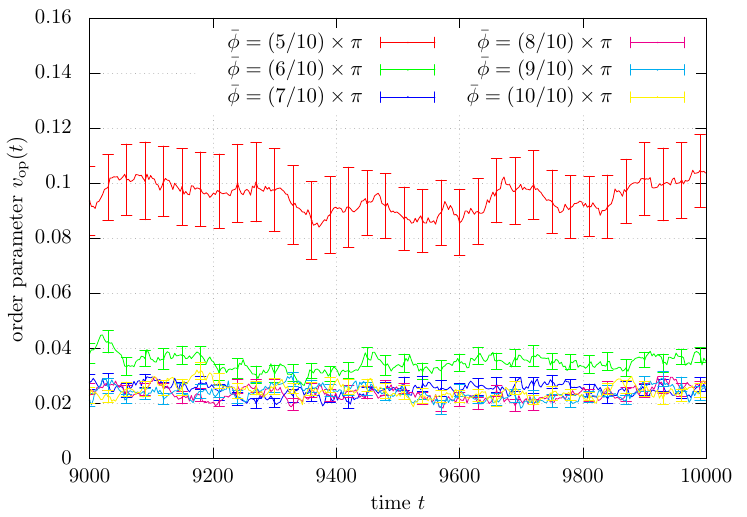}
	\caption{Time evolution of the order parameter, Eq.~\eqref{main_eq_def_order-parameter_Vicsek-model_001_001}, of the $\arctan$ Vicsek model, Eq.~\eqref{main_eq_definition_original-Vicsek-model_001_001}, for several $\bar{\phi}$ in Eq.~\eqref{main_eq_def_phase-factor_001_001}.
		The two panels differ only in the values of $\bar{\phi}$ and the ranges of the horizontal and vertical axes.
		We set $N = 1600$, $L = 8.0$, $v_\mathrm{abs} = 0.01$, $r_\mathrm{V} = 0.1$, $\eta = 0.75$, and $\Delta t = 1.0$. We ran simulations 30 times to compute standard errors.}
	\label{main_fig_gnuplot_time-evolution_order-parameter_001_002}
\end{figure}

The time evolution of the order parameter, Eq.~\eqref{main_eq_def_order-parameter_Vicsek-model_001_001}, of the two models with different additional phases, Eq.~\eqref{main_eq_def_phase-factor_001_001}, is shown in Fig.~\ref{main_fig_gnuplot_time-evolution_order-parameter_001_001}.
We consider a system of size $L \times L$ with periodic boundary conditions and set $N = 1600$, $L = 8.0$, $v_\mathrm{abs} = 0.01$, $r_\mathrm{V} = 0.1$, $\eta = 0.75$, and $\Delta t = 1.0$.
The initial conditions $\{ \bm{x}_i (0) \}_{i=1}^N$ and $\{ \theta_i (0) \}_{i=1}^N$ are sampled independently and uniformly from $[0, L) \times [0, L)$ and $(-\pi, \pi]$, respectively.
The upper panel of Fig.~\ref{main_fig_gnuplot_time-evolution_order-parameter_001_001} shows that the macroscopic behavior of the arctan Vicsek model, Eq.~\eqref{main_eq_definition_original-Vicsek-model_001_001}, strongly depends on the choice of the global phase shift in Eq.~\eqref{main_eq_def_phase-factor_001_001}.
More specifically, when $\bar{\phi} = \pi$ is chosen in Eq.~\eqref{main_eq_def_phase-factor_001_001}, no clear flocking order develops in the arctan Vicsek model, Eq.~\eqref{main_eq_definition_original-Vicsek-model_001_001}, even for the relatively large interaction radius and low noise amplitude used here. This behavior can be attributed to the branch cut of the principal-value representation of the angle.
In contrast, the macroscopic behavior of the arithmetic-mean Vicsek model, Eq.~\eqref{main_eq_mean-angle_angle-Vicsek-model_001_001}, is essentially insensitive to the choice of the global phase shift, as shown in the lower panel of Fig.~\ref{main_fig_gnuplot_time-evolution_order-parameter_001_001}.
This robustness arises because the arithmetic-mean update rule does not involve the principal-value arctangent and therefore does not introduce the same branch-cut dependence.
To further examine the robustness of this observation, we vary $\bar{\phi}$ in Eq.~\eqref{main_eq_def_phase-factor_001_001} and plot the corresponding time evolution of the order parameter in Fig.~\ref{main_fig_gnuplot_time-evolution_order-parameter_001_002}.
We restrict $\bar{\phi}$ to the interval $[0, \pi]$, since the angular variables are represented modulo $2 \pi$ and the results for negative phase shifts are equivalent by symmetry.
Figure~\ref{main_fig_gnuplot_time-evolution_order-parameter_001_002} shows that the order parameter, Eq.~\eqref{main_eq_def_order-parameter_Vicsek-model_001_001}, is strongly suppressed over a broad range of $\bar{\phi}$.
Clear alignment is observed only for relatively small phase shifts, approximately $| \bar{\phi} | \lesssim (2/5) \pi$.
Even in this regime, however, the relaxation time depends strongly on $\bar{\phi}$.
Another difference between the two models is the relaxation time.
Although we do not analyze this issue in detail here, the arithmetic-mean Vicsek model, Eq.~\eqref{main_eq_mean-angle_angle-Vicsek-model_001_001}, reaches a stationary ordered state more rapidly in the parameter regime considered in this study.
This feature may be useful for numerical investigations of certain properties of Vicsek-type models.

\section{Conclusions}

It has been widely assumed that the original Vicsek model possesses $\mathrm{O} (2)$ symmetry and exhibits an order-disorder phase transition associated with the spontaneous breaking of this symmetry.
In this paper, we revisited this premise and demonstrated that $\mathrm{O} (2)$ symmetry is absent in the original angular formulation of the Vicsek model.
Furthermore, we showed numerically that an adaptive choice of the global phase can suppress flocking order even in a regime of low noise amplitude and relatively large interaction radius.
We also showed that, when the mean direction is defined by the arithmetic mean of the angular variables, the corresponding update rule is invariant under global phase shifts.
In this variant, the flocking transition is robust against the adaptive global phase shifts considered here.

\begin{acknowledgments}
	H.M. was supported by JSPS KAKENHI Grants No.~JP25H01499, No.~JP26H01783, and No.~JP26K17043.
\end{acknowledgments}

% \section*{Author contributions}

% \textcolor{red}{(Not finished.)}

% \section*{Competing interests}

% The authors declare no competing financial interests.

% \section*{Data availability}

% We will provide all the numerical codes and plot files upon request.

\appendix

\section{Continuous-time limit of the arithmetic-mean Vicsek model}

Here, we discuss the continuous-time limit of the arithmetic-mean Vicsek model, Eq.~\eqref{main_eq_mean-angle_angle-Vicsek-model_001_001}.
If one is only interested in the onset of flocking order, taking the continuous-time limit may not be necessary; the discrete-time model can even be advantageous because it often requires fewer iterations to reach relaxation.
However, for phases characterized by continuous-time dynamics, such as the chiral phase discussed in Ref.~\cite{Fruchart_001}, such a limit may be necessary.
More generally, physical systems are usually modeled as evolving in continuous time, with discrete-time models regarded as approximations.
This perspective provides a useful criterion for assessing the naturalness of a given discrete-time model.

First, we add a relaxation term to Eq.~\eqref{main_eq_mean-angle_angle-Vicsek-model_001_001} to take the continuous-time limit of the arithmetic-mean Vicsek model, Eq.~\eqref{main_eq_mean-angle_angle-Vicsek-model_001_001}; otherwise, an impulsive force would be present leading to an unnatural model.
By adding a relaxation term to Eq.~\eqref{main_eq_mean-angle_angle-Vicsek-model_001_001}, we obtain
\begin{align}
	\theta_i (t + \Delta t) & = (1 - \alpha) \theta_i (t) + \alpha \langle \theta_j (t) \rangle_{i, r_\mathrm{V}} + \Xi_i (t + \Delta t, t). \label{main_eq_mean-angle_angle-Vicsek-model_relaxation_001_001}
\end{align}
Then, Eq.~\eqref{main_eq_mean-angle_angle-Vicsek-model_relaxation_001_001} can be transformed into
\begin{align}
	 & \frac{\theta_i (t + \Delta t) - \theta_i (t)}{\Delta t} \nonumber                                                                                                                         \\
	 & \quad = \kappa (\langle \theta_j (t) \rangle_{i, r_\mathrm{V}} - \theta_i (t)) + \frac{\Xi_i (t + \Delta t, t)}{\Delta t}                                                                 \\
	 & \quad = \kappa \langle \theta_j (t) - \theta_i (t) \rangle_{i, r_\mathrm{V}} + \frac{\Xi_i (t + \Delta t, t)}{\Delta t}, \label{main_eq_mean-angle_angle-Vicsek-model_relaxation_002_001}
\end{align}
where we have defined $\kappa \coloneqq \frac{\alpha}{\Delta t}$.

In the limit $\Delta t \to 0$, Eq.~\eqref{main_eq_mean-angle_angle-Vicsek-model_relaxation_002_001} becomes
\begin{align}
	\frac{\mathrm{d}}{\mathrm{d} t} \theta_i (t) & = \kappa \langle \theta_j (t) - \theta_i (t) \rangle_{i, r_\mathrm{V}} + \xi_i (t), \label{main_eq_mean-angle_angle-Vicsek-model_continuous-time_001_001}
\end{align}
where $\xi_i (t) \coloneqq \frac{\mathrm{d}}{\mathrm{d}t} \Xi_i (t, \cdot)$.
Assuming $| \theta_j (t) - \theta_i (t) | \ll 1$ for $j \in \{ 1, 2, \dots, N \}$ such that $\| \bm{x}_i (t) - \bm{x}_j (t) \| \le r_\mathrm{V}$, we have $\theta_j (t) - \theta_i (t) \approx \sin (\theta_j (t) - \theta_i (t))$; thus Eq.~\eqref{main_eq_mean-angle_angle-Vicsek-model_continuous-time_001_001} becomes~\cite{Fruchart_001}
\begin{align}
	\frac{\mathrm{d}}{\mathrm{d} t} \theta_i (t) & = \kappa \langle \sin (\theta_j (t) - \theta_i (t)) \rangle_{i, r_\mathrm{V}} + \xi_i (t). \label{main_eq_sin-Vicsek-model_001_001}
\end{align}
Note that Eq.~\eqref{main_eq_sin-Vicsek-model_001_001} does not have a branch cut, unlike the $\arctan$ Vicsek model, Eq.~\eqref{main_eq_definition_original-Vicsek-model_001_001}; consequently, the natural discretization of Eq.~\eqref{main_eq_sin-Vicsek-model_001_001} does not lead to Eq.~\eqref{main_eq_dynamics-angle_original-Vicsek-model_001_001}.

\section{Remark on the rotational symmetry investigated in Ref.~\cite{Baglietto_001}}

Here, we make a remark regarding Ref.~\cite{Baglietto_001}, as they also consider rotational transformations.
However, there is a significant difference between our approach and theirs, and we elaborate on this point in this section.

The rotational symmetry studied in Ref.~\cite{Baglietto_001} involves rotating the positions of the agents with respect to the origin:
\begin{align}
	\bm{x}_i (t) & \mapsto R (\phi (t)) \bm{x}_i (t),
\end{align}
where
\begin{align}
	R (\phi) & \coloneqq
	\begin{bmatrix}
		\cos (\phi) & - \sin (\phi) \\
		\sin (\phi) & \cos (\phi)
	\end{bmatrix}.
\end{align}
Note that $\theta_i (t) \mapsto \theta_i (t) + \phi (t)$ automatically leads to $\bm{v}_i (t) \mapsto R (\phi (t)) \bm{v}_i (t)$.
This definition is problematic because the periodic boundary conditions do not preserve this symmetry; as a result, it fails to capture the intrinsic nature of the $\arctan$ Vicsek model, Eq.~\eqref{main_eq_definition_original-Vicsek-model_001_001}.
Indeed, to the best of our knowledge, no periodic boundary conditions exist that preserve $\mathrm{O} (2)$ symmetry with respect to the origin.

\bibliography{paper_Vicsek_O2-symmetry_999_001}

\end{document}